\documentclass[reqno, centertags]{amsart}
\usepackage{amssymb}
\usepackage{graphicx}

\theoremstyle{definition}

\begin{document}

\title{Quantum Mechanics: 44 admissible questions? \\
-Not only Fapp-}

\author{Michele Caponigro \and Helen Lynn}

\address{Physics Department, University of Camerino, I-62032 Camerino, Italy}
\address{QPT Quantum Philosophy Theories web site www.qpt.org.uk}
\email{michele.caponigro@unicam.it}
\email{helen.lynn@qpt.org.uk}

\date{\today}

\begin{abstract}

The words: determinism, hidden variables, subjectivism,
information, objectivism, informational-theoretic axioms,observers
have some connection with physical reality? What we mean with
"description" of physical reality? When we say that we understand
this reality? Certain parameters:position, velocity are
sufficient? We will focus only to conceptual considerations
regarding the relation between the "questions" and the relative
"answers" in general and specifically in quantum mechanics. It is
usually believed that the answers are more important of the
questions, for this reason we can read many answers everywhere and
in different field of knowledge. We need to add and clarify some
things: (i) usually an answers require a question, (ii) but, as we
know, their relation is not so simple and immediate, (iii) For
instance: a)an epistemic questions give us ontic answers? b)the
answer has a connection with the question and vice versa? c)we
could to infer a question starting from an answer? d)there are
answers without questions? These answers could be in some
framework considered as ontic answers? The relative scientific
works are the same time ontic? Speaking of quantum mechanics we
see around many answers in the meantime we do not see the
correspondent questions, these answers seem completely
independent, and this seem a right road, the road of the
independent nature unlinked from human thoughts. We retain instead
that questions can affect the possible answers. Exist "something"
before the question?

\end{abstract}

\maketitle

\section{A good news: a brief paper without answers}

\begin{enumerate}
\item Quantum theory describe physical reality?
\item Wavefunction is only a mathematical expression for evaluating probabilities?
\item The wavefunction is not an objective entity?
\item The time dependence of the wavefunction does not represent the
evolution of a physical system?
\item What the collapse of the wave function mean?
\item What causes it?
\item What is a measurement?
\item How does physics to describe reality?
\item The interpretation is merely philosophical bias, and
therefore no part of physics.
\item The wave function is said to refer exclusively to a human
mind and not any physical system external to that mind?
\item Theoretical terms are never directly observable?
\item Determinism has a connection with the physical reality?
\item Why we can ask about the problem of hidden variables? It is
admissible this question?
\item Is the above question supported by some tacit assumption?
Before the question, already we have in mind a picture of the
physical reality?
\item We understand better the physical reality with hidden variables?
\item It is possible have the hidden of the hidden variables?In that case the nature will be more intelligible?
\item It is admissible the link between determinism and the hidden variables?
\item Which is the nature of the Probability?
\item Why we introduce the notion of probability?
\item What is the probability of a single object?
\item Probability is merely a collective term?
\item How we introduce the properties of an object?
\item The properties of a single object are the same of an ensemble (of the same) objects?
\item We need to assume the existence of the observers?
\item How is the information content of a system quantified?
\item How is information transferred?
\item What is the physical status of information?
\item What role, if any, can an information-theoretic analysis of a physical phenomenon
      play in an explanation of a physical phenomenon?
\item The information is something of physical(ontological problem)?
\item Any theorem is valid only for a fixed system of axioms? Why?
\item It is possible to have another axiomatic of quantum mechanics?
\item Where the GRW-collapse occur?
\item The standard view: an ontic measurement of an epistemic reality?
\item Classical ensemble probability could not work for ensembles of quantum systems?
\item Which is the origin of the randomness?
\item Which is the statute of quantum realism?
\item It is admissible an holistic nature of wavefunction?
\item It is possible a distinction within a quantum state between
ontic and epistemic elements?
\item Is the brain a quantum system?
\item Have quantum mechanical isolated systems a physical meaning?
\item The geometry could eliminate the observers?
\item Quantum information exist?
\item Probability require indistinguishability?
\item Physical Systems vs. Conceptual Systems?
\end{enumerate}

\section{Conclusion}
The pretext of 44 questions is only to affirm the importance of
the "questions" in our research, probably some questions listed
above will be: a)inadmissible b) of metaphysical nature c)with
simple answers.

\end{document}